\documentclass[texture,10pt]{article}
\def\beq{\begin{equation}}
\def\eeq{\end{equation}}
\def\ben{\begin{eqnarray}}
\def\een{\end{eqnarray}}
\def\bea{\begin{array}}
\def\eea{\end{array}}

\begin{document}

\baselineskip=18pt
\begin{flushright}
Prairie View A \& M, HEP-12-98\\

({\bf Draft})
\date{today}

\end{flushright}

\vskip.1in

\begin{center}
{\Large\bf Simple unitarity relations among charged current coupling 
constants}
\vskip .3in

{\bf Dan-Di Wu}\footnote{ $~$E-mail:  wu@hp75.pvamu.edu or
danwu@flash.net }
\vskip.1in

{\sl HEP, Prairie View A\&M University, Prairie View, TX 77446-0932, USA}
\end{center}
\vskip.5in

{\sl  I suggest two new unitarity tests for the quark charged  current 
coupling constants which are
$({\rm sin}\gamma/{\rm sin}\alpha)=\zeta$, and 
$\frac{{\rm sin}2\gamma}{{\rm sin}2\alpha}=-\frac{(1-\zeta 
{\rm cos}\beta)}
{{\rm cos}\beta-\zeta}$
where $\zeta$ is a CKM parameter defined in the text.
These unitarity identities do not suffer some of the multi-value ambiguities. 
Related problems are discussed. The sign of $\Delta m_{_{B_d}}$
should satisfy the condition $\Delta m {\rm cos}2\beta < 0$.
}

\vskip.4in

The accelerators and detectors for the measurement of CP violation in the B-
meson system are close to completion at both SLAC and KEK. Preparation for
 data acquisition and analysis is intensified\cite{SLAC}. 
The question of focus is still how to efficiently use
limited resources (e.g. the limited data that will be available) 
to test the standard model (SM), in particular, 
the unitarity of the charged current coupling constants,
which in the SM are grouped into the Cabibbo-Kobayashi-Maskawa (CKM) 
matrix\cite{CKM}. In this note I would 
 present two unitarity relations for discussion.
\vskip.08in
The first new unitarity relation I suggest is
\beq
{\rm sin}\gamma/{\rm sin}\alpha=\zeta
\eeq
where $\alpha, \gamma$ are the  unitarity angles
discussed widely in the literature and $\zeta$ is a
CKM parameter that will be defined the next moment.
I will also suggest another
 unitarity raltion in which only ${\rm sin}2\alpha$ and ${\rm sin}2\gamma$
appear in the left-hand side of the equation.  
\vskip.08in
In the standard model, the CKM Matrix
\ben
V_{CKM}=\left(\bea{ccc}
V_{ud}&V_{us}&V_{ub}\\
V_{cd}&V_{cs}&V_{cb}\\
V_{td}&V_{ts}&V_{tb}\eea\right)
\een
is believed to be the only coupling constant matrix 
that couple quarks of different flavors to positively 
charged W-bosons. It therefore plays an exclusive role
in the standard model CP violation phenomenology, since there
is not any other charged boson in the standard model, 
neither  a vector nor  a scalar meson. This 
matrix is supposed to be unitary, because of its gauge 
interaction origin. Beyond the standard 
model there may be new contributions from additional couplings
in  some new physics to the 
processes which exist in the SM, including flavor changing neutral
current processes. These new couplings cause an apparent violation of unitarity 
of the measured effective coupling constants. Unitarity
test is therefore important to discovering of possible physics
beyond the standard model.
\vskip.08in
A clever way of testing unitarity is to work on a
unitarity triangle, which is defined by one of the orthogonality
conditions. A triangle tightly related to $B_d$ mixing and decay
is defined by the following unitarity condition
\beq
V_{ud}V_{ub}^*+V_{cd}V_{cb}^*+V_{td}V_{tb}^*=0.
\eeq
Regarding the three terms in this equation as three sides,
the three outer-angels are defined as 
\beq
\alpha={\rm arg}(B/C), \beta={\rm arg}(C/A), \gamma={\rm arg}(A/B), 
\eeq
where $A, B, C$ are defined as
\beq
A=V_{cd}V_{cb}^*, B=V_{ud}V_{ub}^*, C=V_{td}V_{tb}^*.
\eeq
By definition,
\beq
\alpha+\beta+\gamma=2\pi
\eeq
no matter  $(A+B+C)$ vanishes or not. 
However, some bizarre 
new physics may 
cause the measured angles dissatisfy Eq (6), (e.g. if the measured angles
break the biting-tail relations   in (4)) although in terms
of unitarity test, Eq (6) is not very useful. What is  special of Eq (1)
is that it will not stand unless the measured effective
parameters are those derived from a unitary CKM matrix.
\vskip.08in
The relation  Eq(1) can be easily proved using the matrix convention
of Chen and Wu\cite{Chen}, which reads
\ben
V_{CKM}=\left(\bea{ccc}
1-\frac{1}{2}\lambda^2&\lambda
&A\lambda^3(e^{-i\beta}-\zeta)\\
&&\\
-\lambda+A^2\zeta\lambda^5e^{-i\beta}&1-\frac{1}{2}
\lambda^2&A\lambda^2e^{-i\beta}\\
&&\\
A\zeta\lambda^3&-A\lambda^2e^{i\beta}&1\eea\right),
\een
where $\lambda$ and $A$ are the parameters a la Wolfenstein\cite{Wol}.
The currently best fit values for the two parameters are respectively 
$\lambda=0.220\pm0.002$ and $A=0.85\pm0.07$.
$\beta$ is exactly the $\beta$-angle discussed above.
$\zeta$ in Eq (1) is also defined here. By definition,
$\lambda, A, \zeta >0$ is required. 
This expression of the CKM matrix is accurate to the third oreder
of $\lambda$  (i.e. $\lambda^3)$), which is good enough for most practical
usages.\footnote{The expression accurate to the fifth order of $\lambda$ 
is
$$
V_{CKM}=\left(\bea{ccc}
1-\frac{1}{2}\lambda^2-\frac{1}{8}\lambda^4&
\lambda
&A\lambda^3(e^{-i\beta}-\zeta+\frac{1}{2}\lambda^2e^{-i\beta})\\

&&\\
-\lambda+A^2\lambda^5(\zeta e^{-i\beta} -\frac{1}{2})
&1-\frac{1}{2}
\lambda^2-(\frac{1}{8}+\frac{1}{2}A^2)\lambda^4
&A\lambda^2(e^{-i\beta}+\zeta \lambda^2+\frac{1}{2}\lambda^2e^{-i\beta})\\

&&\\
A\zeta\lambda^3(1+\frac{1}{2}\lambda^2)
&-A\lambda^2e^{i\beta}(1+\frac{1}{2}\lambda^2)
&1-A^2\lambda^4\eea\right)
$$}
\vskip.08in

The three sides of the master triangle discussed above
are defined respectively,
after a scaling
\beq
A=-e^{i\beta}, \hskip.2in B=e^{i\beta}-\zeta, \hskip.2in C=\zeta.
\eeq
Combining (8) with (4), we have 
\beq
{\rm sin}\alpha=\frac{{\rm sin}\beta}{\sqrt{1+\zeta^2-2\zeta{\rm cos}\beta}},
\hskip.3in
{\rm sin}\gamma=\frac{\zeta{\rm sin}\beta}{\sqrt{1+\zeta^2-2\zeta{\rm 
cos}\beta}}.
\eeq
Eq (1) therefore follows.
\vskip.08in
We emphasize that Eq (1) comes from the unitarity  of the CKM matrix
expressed by $A+B+C=0$. It does not
predict the value of any parameters, unless two of the three parameters
in the equation are measured. As mentioned before, Eq (1) relates two physical
parameters
to a third measurable, which is a CKM parameter. (We will come back to the
question of how
to measure $\zeta$ later on.) However,
there are some ambiguities in practice, because what
seem  likely to be measured in relevant CP violating processes
are ${\rm sin}2\alpha, {\rm sin}2\gamma$ etc. 
To calculate
${\rm sin}\alpha$ from measured ${\rm sin}2\alpha$, one obtains two pairs 
of solutions, each pair involve two solutions with opposite signs.
It is therefore useful to provide also another unitarity relation
\beq
\frac{{\rm sin}2\gamma}{{\rm sin}2\alpha}=-\frac{(1-\zeta {\rm cos}\beta)}
{{\rm cos}\beta-\zeta}
\eeq
\vskip.08in
 The value of cos$\beta$ in the right-hand side of Eq (10)
 will be uniquely defined, if both sin$2\beta$ and 
sin$\beta$ are directly measured
from CP violation effects. It is well known that
sin$2\beta$ is exactly the asymmetry in the $B_d\rightarrow J/\psi K_S$ 
decay which was first examined by Bigi, Carter and Sanda \cite{Sanda},
\beq
a_{B_d\rightarrow J/\psi K_S}=sin2\beta.
\eeq
This quantity will be measured cool for both CP violation study
and the CKM parametrization. Its signal at the two B-factories 
will be very clean. 
The value of sin$\beta$ can be obtained by the measurement of like-charged lepton
asymmetry\cite{Wu1}, once $\zeta$ is measured
\beq
a_{ll}=2{\rm Im}(\Gamma_{12}/M_{12})_{B_d}, 
\eeq
where $a_{ll}$ is defined as 
$$a_{ll}=\frac{N(B^0\bar B^0\rightarrow l^+l^+)-N(B^0\bar B^0\rightarrow 
l^-l^-)}
{N(B^0\bar B^0\rightarrow l^+l^+)+N(B^0\bar B^0\rightarrow l^-l^-)}.$$
The right-hand side of Eq (12) is
\beq
2{\rm Im}(\Gamma_{12}/M_{12})_{B_d}=\frac{8\pi}{F_{box}(m_t^2/m_W^2)}
\frac{m_c^2}{m_t^2}\frac{{\rm sin}\beta}{\zeta},
\eeq
where $F_{box}(x)$ is a box diagram function defined by Inami and 
Lim\cite{Inami}.
\vskip.08in
There is a draw back in Eq (10) concerning the accuracy of the denominator in 
its right-hand side. If, say,
${\rm cos}\beta$ is measured of value 0.70$\pm 0.07$ with 10\% of accuracy,
and $\zeta$ is measured of value 0.50$\pm 0.10$ with 20\% of accuracy, then 
the denominator  of the right-hand side of this formula will be 0.20
$\pm 0.12$ which is 60\% 
uncertain. Such error enlargement may become worse when the calculations
involved are more complicated.
We will have a very bad luck in using this formula for unitarity test,
if a situation similar to or worse than this happens.
There are many other methods to measure  sin2$\beta$ and sin$\beta$,
see the BaBar physics book\cite{SLAC}.
\vskip.08in
It might be a good place to clarify the relation between $a_{ll}$ and 
$\epsilon_{_{B_d}}$. The question is 
$$
{\rm Re}\epsilon_{_{B_d}} ?= \frac{1}{4}a_{ll}.
$$ 
We put a question mark here to
make an alarm.
To show its invalidity, one finds, from the eigenequations
for $\left(\bea{c}(1+\epsilon)\\ (1-\epsilon)\eea\right)$, 
$$
\frac{(1+\epsilon)^2}{(1-\epsilon)^2}=\frac{M_{12}-\frac{i}{2}\Gamma_{12}}
{M_{12}^*-\frac{i}{2}\Gamma_{12}^*}.
$$
Defining $M_{12}=-|M_{12}|e^{i\theta}$ ($\theta$ is the phase of the mixing 
mass), 
$\sigma=\frac{1}{4}{\rm Im}(\Gamma_{12}/
M_{12})$ ($\sigma$ also equals to $\frac{1}{2}\langle B_+|B_-\rangle$ and is 
therefore called
the overlap of the two CP eigenstates)  
and $R=-{\rm Re}(\Gamma_{12}/M_{12})$, and noting that $\sigma$ 
and $R$ are very small for the neutral $B$ meson systems, one finds,
\beq
\epsilon_{_{B_d}}=\frac{(1+\sigma)-e^{i\theta}(1-\sigma)}
{(1+\sigma)+e^{i\theta}(1-\sigma)}.
\eeq
 The value of $\theta=$ arg$(V_{tb}
V_{td}^*)^2$ is CKM convention dependent. 
For $1-{\rm cos}^2\theta >> \sigma\sim 10^{-3}-10^{-4}$, 
one obtains\cite{Wu}
$$
{\rm Re}\epsilon_{_{B_d}}=\frac{2\sigma}
{1+{\rm cos}\theta+\sigma^2(1-{\rm cos}\theta)}.
$$
The $\theta$ and $\epsilon_{_{B_d}}$ values are
 listed in the following table for different
CKM parametrizations.
$$
\bea{ccc}
\hline\\
{\rm CKM parametrization}& \theta&{\rm Re}\epsilon_{_{B_d}}\\
&&\\
\hline
{\rm Chen-Wu}& 0& \sigma\\
{\rm Wolfenstein}& \,\,+{\rm arctan}(\frac{\eta}{1-\rho})\,\,&
\frac{2\sigma}{(1+{\rm cos}\theta)+\sigma^2
(1-{\rm cos}\theta)}\\
\,\,{\rm Kobayashi-Maskawa}\,\, &  \delta&
\,\,\frac{2\sigma}
{1+{\rm cos}\delta+\sigma^2(1-{\rm cos}\delta)}\,\,\\
\hline\\
\eea
$$

 In the 
Wolfenstein parametrization, sin$2\alpha$ and sin$2\beta$ are respectively\cite{SLAC}:
\beq
{\rm sin}2\alpha=\frac{2\bar\eta[\bar\eta^2+\bar\rho(\bar\rho-1)]}
{[\bar\eta^2+(\bar\rho-1)^2][\bar\eta^2+\bar\rho^2]},\hskip.2in
{\rm sin}2\beta=\frac{-2\bar\eta(\bar\rho-1)]}{\bar\eta^2+(\bar\rho-1)^2},
\eeq
where
$\bar\eta=\eta(1-\lambda^2/2)$, $ \bar\rho=\rho(1-\lambda^2/2).$ Obviousely,
it is very difficult to collect information about unitarity  from the
comparison of these two equations and related measurements.  
\vskip.08in
The value of $\zeta$   can  be acquired from 
 the mixing mass of the $B_d-\bar B_d$ system  
\beq
|\Delta m_{B_d}|=\frac{G_F^2}{6\pi^2}\eta (Bf_B^2)m_t^2F_{box}(m_t^2/m_W^2)
A^2\lambda^6\zeta^2.
\eeq
where $\eta$ is a QCD correction factor, B is the vacuum insertion,
or box constant, and $f_B$ is the $B_u$ to $\mu \bar\nu$ decay
constant. From this formula the current
likelihood of $\zeta$ is $0.73\pm 0.29$, which is very much dependent 
on the accuracy of the calculation of $\eta (Bf_B^2)$.
${\rm cos}\beta$ can then be
calculated from the ($B \rightarrow u+x)$ to ($B \rightarrow c+x)$ ratio
\beq
\frac{1}{\lambda}\left|\frac{V_{ub}}{V_{cb}}\right|=\sqrt{1+\zeta^2-2\zeta
{\rm cos}\beta}.
\eeq
From this formula, cos$\beta$ is obtained, cos$\beta=0.9\pm 0.4$. 
The uncertainty of cos$\beta$ here is mainly from the uncertainty of
$\zeta$ itself. The right-hand side of Eq (16) can also be written as
$\sqrt{(\zeta-{\rm cos}\beta)^2+{\rm sin}^2\beta}$, from which 
one finds that
$$|{\rm sin}\beta)|\le 0.36\pm 0.07.$$
There is not any error enlargement mechanism involved in Eq (16)
because of the smallness of the ($b$ to $u$)/($b$ to $c$) ratio in
the left-hand side of this equation,
which strongly suggests 
that cos$\beta$ must be positive.
\vskip.08in
It becomes apparent with the Chen-Wu matrix that the two eigenstates
\beq
B_+=\sqrt{\frac{1}{2}}\left[(1+\epsilon)|B_d\rangle + 
(1-\epsilon)|\bar B_d\rangle\right], \hskip.2in 
B_-=\sqrt{\frac{1}{2}}\left[(1+\epsilon)|B_d\rangle - 
(1-\epsilon)|\bar B_d\rangle\right]
\eeq
are almost pure CP eigenstates, because $\epsilon$ here is extremely small.
 This should be true even if one uses
other CKM matrix conventions in which $\epsilon$ is large, 
because CP violation in mixing, $a_{ll}$
is very small anyway. An interesting question one may ask is
what  the sign of the mass difference $\Delta m$ 
of the two eigenstates is, where $\Delta m$ is defined as
\beq
\Delta m = m_+-m_-.
\eeq
The idea was that possible large CP violation in the decay amplitudes
might cause the CP even $B_d$ eigenstate to mainly decay into CP odd final states
 and the CP
odd eigenstate into CP even final states. Consequently, the width of 
$B_+$ became less than that of $B_-$. 
From 
\beq
\Delta m \Delta \gamma=4{\rm Re}(M_{12}\Gamma_{12}),
\eeq
one would then find that $\Delta m > 0$, where $\Delta \gamma$
was the corresponding width difference.  Indeed, the leading CKM term in 
$\Gamma_{12}$ was also $(V_{tb}V_{td}^*)^2$ as calculated by 
Hagelin\cite{Hagelin} , so the right-hand side
of Eq (20) was negative. One knows that $\Delta m_K$ is negative, but
now  $\Delta m_{_{B_d}}$
could be positive\cite{Bruce}.
\vskip.08in
With the Chen-Wu matrix, the discussion is simple, 
because of the smallness of $\epsilon_{_{B_d}}$ in this convention.
The criteria\cite{Wu2} for a judgement is 
\beq
\Delta m {\rm Re}
\Delta^2_{us}<0,
\eeq
where 
\beq 
\Delta_{us}=V_{cb}
V_{td}V^*_{cd}V^*_{tb},
\eeq
is a relevant quartet rephasing invariant of the CKM matrix. 
This can be further simplified as
\beq
\Delta m_{_{B_d}} {\rm cos}2\beta < 0.
\eeq
That means the ``anomalous" sign for $\Delta m_{_{B_d}} $ will appear, if 
$\beta $ is in between $\pi/4$ and $3\pi/4$. 
\vskip.08in

This work is in part supported by the Department of Energy and in part by 
the Center for Applied Radiation Research (CARR) at PVAMU.
\vskip.4in


\end{document}